\documentclass[aps,prl,reprint,superscriptaddress]{revtex4-2}

\usepackage{amsmath}
\usepackage{amsfonts}
\usepackage{comment}
\usepackage{graphicx}
\newcommand{\Tr}{\mathrm{Tr}}

\newcommand{\vev}[1]{\langle #1 \rangle}
\usepackage{xcolor}
\usepackage{color}

\begin{document}


\title{Dynamics of the Density Cube}

\author{Nabin Bhatta}
\email{nabinb@vt.edu}

\author{Djordje Minic}
\email{dminic@vt.edu}

\author{Tatsu Takeuchi}
\email{takeuchi@vt.edu}

\affiliation{Center for Neutrino Physics, Department of Physics, Virginia Tech, Blacksburg VA 24061, USA}

\date{\today}

\begin{abstract}
Density cube theory extends the canonical quantum density matrix $\rho_{ij}$ with the addition of
an extra index to $\rho_{ijk}$. The elements of the density cube with two different indices, $\rho_{iij}$ and $\rho_{ijj}$, correspond to the
real and imaginary parts of the off-diagonal element $\rho_{ij}$ of the density matrix and describe double-path interference, 
while those with three different indices describe non-canonical triple-path interference.
In this letter, we propose an equation of motion for the density cube, obtained
from the quantization of ternary Nambu dynamics, and find that pairs of triple-path interferences
oscillate into each other.
\end{abstract}

\maketitle
\noindent
{\it Introduction:} 
One of the features that distinguishes Quantum Mechanics (QM) from Classical Mechanics (CM) is the existence of interference between paths. However, as pointed out by Sorkin \cite{Sorkin:1994dt}, quantum interference is limited to pairwise path interference due to the Born rule.
To introduce triple-path interference and higher one must go beyond the canonical QM formulation, and any experimental detection of such interferences would signal new physics beyond canonical QM.

Experiments have limited the size of triple-path interference relative to that of double-path interference to order $10^{-3}$ \cite{Sinha:2010, Park:2012, Sawant2014, Kauten_2017, Cotter2017, Sadana2022, Namdar2023}.
On the theoretical side,
works have mostly focused on the operational questions, investigating whether higher-order interference provides advantage in certain information processing tasks~\cite{lee2016deriving}, or whether it implies correlations stronger than canonical QM~\cite{Niestegge:2013cua}. It has been also investigated whether the absence of third-order interference can be used as an axiom to derive the formalism of QM~\cite{Barnum_2014}. 

A formalism exhibiting triple-path interference is the density cube theory proposed by Daki\'{c} et al. \cite{Dakic_2014} 
\footnote{Another extension of QM containing triple-path interference is quartic quantum theory developed in Zyczkowski \cite{zyczkowski2008quartic}. For a comparative study of density cube theory and quartic quantum theory, see \cite{Lee_2016}.}.
There, the two-index density matrix $\rho_{ij}$ of canonical QM is 
extended to the three-index density cube $\rho_{ijk}$,
where the subscripts are Hilbert space indices.
The inner product between two cubes $A_{ijk}$ and $B_{ijk}$ is defined as
\begin{equation}
(A,B) = \sum_{ijk} A_{ijk}^*B_{ijk}\;,
\end{equation}
and the expectation value of observable $X_{ijk}$ 
for the density cube $\rho_{ijk}$ is given by
%
$\vev{X} = (\rho,X)$.
%
To ensure the reality of this expectation value, we 
demand both the density cube $\rho_{ijk}$ and
the observable $X_{ijk}$ to be ``hermitian,'' 
where we define a cube $A_{ijk}$ to be hermitian when
for any permutation $\pi$ of the three indices it satisfies
\begin{equation}
A_{\pi(ijk)} 
=
\begin{cases}
A_{ijk}   & \mbox{if $\pi$ is an even permutation} \\
A_{ijk}^* & \mbox{if $\pi$ is an odd permutation} 
\end{cases}
\label{CubeHermiticity1}
\end{equation}
This guarantees that any transposition of the indices will complex conjugate the element so that
$(\rho,X)^* = (\rho,X)$,
and imposes the following conditions on the elements of a hermitian cube:
\begin{equation}
\begin{array}{l}
A_{iii}\in\mathbb{R}\;,\quad
A_{iij} = A_{iji} = A_{jii} \in \mathbb{R}\;,\\
A_{ijk} = A_{jki} = A_{kij} = A_{ikj}^* = A_{jik}^* = A_{kji}^*\;.
\end{array}
\label{CubeHermiticity2}
\end{equation}
These are the analogs of the conditions $A_{ii}\in\mathbb{R}$, $A_{ij}=A_{ji}^*$ for hermitian matrices.
Recall that for an $N$-level system a generic hermitian matrix $A_{ij}$ has $N^2$ real parameters, of which $N$ are the
diagonal elements $A_{ii}$, and $N(N-1)$ are the real and imaginary parts of the off-diagonal elements $A_{ij}$.
In constrast, a generic hermitian cube $A_{ijk}$ 
for an $N$-level system has $N(N^2+2)/3$ real parameters, 
of which $N$ are the diagonal elements $A_{iii}$, $N(N-1)$ are the elements with two identical indices $A_{iij}$ and $A_{ijj}$, and $N(N-1)(N-2)/3$ are the real and imaginary parts of elements with three different indices $A_{ijk}$.

Define the ternary delta as
\begin{equation}
\delta_{ijk} = \begin{cases}
1 & \mbox{when $i=j=k$} \\
0 & \mbox{otherwise}
\end{cases}    
\end{equation}
and the ``trace'' of a cube as
\begin{equation}
\Tr\,A = (\delta,A) = \sum_{i}A_{iii}\;.
\end{equation}
In \cite{Dakic_2014}, the ``diagonal'' element $\rho_{iii}$ 
of the density cube is identified as the probability that the 
outcome associated with the
$i$th path is obtained, so they are normalized to
\begin{equation}
\Tr\,\rho = (\delta,\rho) = \sum_{i}\rho_{iii} = 1\;.
\label{TotalProb}
\end{equation}
%
Note that since we assume that the indices of the density cube are Hilbert space indices,
this condition cannot be maintained under changes of basis of the Hilbert space:
For the density matrix, trace preservation under basis change is tied to the fact that trace equals the sum of the matrix's eigenvalues. For cubes, there is no analogous relation.
Thus, this probability interpretation requires us to work in a fixed Hilbert space basis.
\footnote{%
We note that in the operational framework\cite{Hardy2001quantum} in which the density cube theory of \cite{Dakic_2014} is formulated, 
the primitive elements of the theory are preparation of states (density cubes in this case), transformation of those states, and measurement of outcomes. Since the framework is completely general, it includes theories where states may not be interpretable as residing in an underlying Hilbert space.
That is, the indices of the density cube may not necessarily be Hilbert space indices.
We will not consider this possibility in this paper.
}

The elements of the density cube with two distinct indices $\rho_{iij}$ and $\rho_{ijj}$ describe the interference between the $i$th and $j$th paths. These are specified by $N(N-1)$ real parameters, matching
the number for the density matrix.
Indeed, the diagonal and two-district-index elements of the 
density cube can be mapped to the elements of the density matrix $\rho_{ij}$ via \cite{Dakic_2014}
\begin{equation}
\rho_{iii}=\rho_{ii},\;\;
\rho_{iij}=\sqrt{\frac{2}{3}}\mathrm{Re}\,\rho_{ij},\;\;
\rho_{ijj}=\sqrt{\frac{2}{3}}\mathrm{Im}\,\rho_{ij},\;\;(i<j).
\label{MatrixCubeCorrespondence}
\end{equation}
The elements of the density cube with three distinct indices, $\rho_{ijk}$, are new degrees of freedom. 
They encode the triple-path interference of the $i$th, $j$th, and $k$th paths.

To describe the dynamics of the density cube, \cite{Dakic_2014}
introduces an analog of a unitary transformation to map 
a hermitian trace-one density cube to another.
However, an equation of motion which specifies the transformation
for an arbitrary time interval was lacking.
In this letter, we propose a possible formalism for this dynamics.
The starting point of our work is the generalized Hamiltonian dynamics of Nambu \cite{Nambu:1973qe, yoneya2026}.

\noindent
\textit{Ternary Nambu Dynamics and its quantization:}
In 1973 Nambu introduced an ingenious generalization of canonical Hamiltonian dynamics \cite{Nambu:1973qe}. 
The classical equation of motion based on the Poisson bracket 
is generalized to 
\begin{equation}
\dfrac{d}{dt}\rho = \{\rho,H_1,H_2\}\;,   
\label{NambuCM}
\end{equation}
where the Nambu bracket on the right-hand side is defined as 
\begin{equation}
\{A,B,C\} = \sum_{ijk}\varepsilon_{ijk}\,
\dfrac{\partial A}{\partial q_i}
\dfrac{\partial B}{\partial q_j}
\dfrac{\partial C}{\partial q_k}
\;.
\label{NambuBracket}
\end{equation}
This bracket satisfies the fundamental identity
\cite{Hoppe:1996xp, Takhtajan:1993vr}
\begin{eqnarray}
& &
\{\{A,B,C\},D,E\} = 
\{\{A,D,E\},B,C\}
\cr
& & \qquad
+ \{A,\{B,D,E\},C\} + \{A,B,\{C,D,E\}\},
\quad
\label{NambuFundamental}
\end{eqnarray}
which implies the Leibniz rule for the operator $\{*,H_1,H_2\}$, justifying its identification with the time derivative.

The Nambu bracket generalizes the antisymmetric structure of the Poisson bracket.
Area-preserving diffeomorphisms generated by the Poisson bracket are generalized to volume-preserving diffeomorphisms generated by the Nambu bracket. 
A prime example of Nambu dynamics is
the free asymmetric top 
with $\rho$, $H_1$, and $H_2$ respectively corresponding to the top's angular momentum, total kinetic energy, and total angular momentum squared in the body-fixed frame \cite{Nambu:1973qe, Takhtajan:1993vr}.
For a recent review, see \cite{yoneya2026}.

Given the well-known correspondence between classical and quantum equations of motion~\cite{Dirac1925}, 
Nambu also posed the natural question:
what is the ``quantum'' analog of his generalized Hamiltonian dynamics? \cite{Nambu:1973qe}. 
Many authors have tackled this question 
\cite{Takhtajan:1993vr,Hoppe:1996xp,Dito:1996hn, Dito:1996xr, Awata:1999dz,Kawamura:2003cw,Kawamura:2005ic,Curtright:2002fd,Curtright:2008jj,Curtright:2009qf, Yoneya:2021lyf}. If we are to describe the ``quantum'' system
with a density cube,
the natural solution would be to assume that the quantum version of Nambu dynamics is given formally by
\begin{equation}
\dfrac{d}{dt}\rho = i[\rho,H_1,H_2]\;,
\label{NambuQM}
\end{equation}
where $\rho$, $H_1$, and $H_2$, are all hermitian cubes,
and $[A,B,C]$ is the ternary commutator of cubes, which we will call the ``termutator'' for short, defined as
\begin{equation}
[A,B,C] = ABC + BCA + CAB - ACB - BAC - CBA\;.
\end{equation}
The termutator maintains the antisymmetry of the Nambu bracket so that if any two entries are the same, it is automatically zero, ensuring that the two Hamiltonians are conserved since $[H_1,H_1,H_2]=[H_2,H_1,H_2]=0$.
However, 
the consistency of Eq.~\eqref{NambuQM} hinges
on how the ternary product of three cubes $ABC$
is defined, and this is where the various works in the literature differ.

There are several requirements that the ternary product should satisfy.
First, the most obvious condition is that
it must be tri-linear in the three cubes
that constitute the product.
Second, the resulting termutator must be such that the right-hand side of Eq.~\eqref{NambuQM} is hermitian 
so that the hermiticity of the density cube is maintained under time evolution.
Third, the termutator of hermitian cubes must
be traceless so that the total probability, Eq.~\eqref{TotalProb}, 
is also conserved.
Fourth, the resulting termutator must satisfy
the fundamental identity 
\begin{eqnarray}
& & [[A,B,C],D,E] = [[A,D,E],B,C] \cr 
& & \qquad
+ [A,[B,D,E],C] + [A,B,[C,D,E]]\;,
\label{FundamentalNambuCommutator}    
\end{eqnarray}
so that the operator $i[*,H_1,H_2]$ satisfies the Leibniz rule, just like the Nambu bracket.

This final condition turns out to be
the most difficult to realize
\cite{Awata:1999dz,Takhtajan:1993vr,Dito:1996hn,Dito:1996xr,Abramov2009AlgebrasWT,Aader2025,Blumenhagen:2025nla} since it requires the
ternary product to be ``associative'' in the sense that
\begin{equation}
(ABC)DE = A(BCD)E = AB(CDE) = A(DCB)E\;,
\label{Associativity}
\end{equation}
that is, the hermitian cubes must simultaneously be
totally associative ternary algebras of both
the first and second kinds \cite{Abramov2009AlgebrasWT}.
While it is not impossible to construct such a ternary product, the definition turns out to be quite complicated \cite{Awata:1999dz}.
An alternative is to start from a fairly simple
definition of the ternary product and restrict our attention to a subalgebra of the hermitian cubes over which the fundamental identity is satisfied.

\noindent
\textit{The Kawamura Product:}
To this end, we adopt the definition of the ternary product introduced by Kawamura \cite{Kawamura:2003cw},
\begin{equation}
(ABC)_{ijk} \;=\; \sum_\ell A_{ij\ell}B_{i \ell k}C_{\ell jk}\;,  
\label{KawamuraProduct}
\end{equation}
where repeated indices are not automatically summed, that is, $i$, $j$, and $k$ are not dummy indices.
Note that this is an obvious analog of the product of two matrices $(AB)_{ij} =\sum_\ell A_{i\ell}B_{\ell j}$, and can be readily extended
further to the product of $n$ hypercubes with
$n$ indices each. (See Appendix A of \cite{Kawamura:2005ic}.)

Under permutations of the indices, the ternary product of three hermitian cubes transforms as
\begin{eqnarray}
(ABC)_{jik}
& = & (ACB)_{ijk}^* \;,\cr
(ABC)_{kji}
& = & (CBA)_{ijk}^* \;,\cr
(ABC)_{ikj}
& = & (BAC)_{ijk}^* \;,\cr
(ABC)_{jki}
& = & (BCA)_{ijk} \;,\cr
(ABC)_{kij}
& = & (CAB)_{ijk} \;,
\end{eqnarray}
which ensures that $i[A,B,C]$ will be hermitian if $A$, $B$, and $C$ are. 
The trace of the ternary product of three cubes is
\begin{equation}
\Tr(ABC)
= (\delta,ABC) 
= \sum_{ij}A_{iij}B_{iji}C_{jii}
\;,
\end{equation}
analogous to $\Tr(AB)=\sum_{ij}A_{ij}B_{ji}$ for matrices.
If the cubes $A$, $B$, and $C$ are all hermitian, then their elements that appear in the above expression are all real and invariant
under any permutation of the indices.
This implies that the trace of the ternary product of three hermitian cubes does not depend on the ordering of the cubes:
\begin{eqnarray}
\lefteqn{\Tr(ABC)=\Tr(BCA)=\Tr(CAB)}\cr
& = & \Tr(ACB)=\Tr(BAC)=\Tr(CBA).
\end{eqnarray}
This is the analog of the relation
$\Tr(AB)=\Tr(BA)$ for matrices though in the case of matrices $A$ and $B$ do not have to be hermitian for this relation to hold.
Therefore, $\Tr[A,B,C] = 0$.    

\noindent
\textit{The Fundamental Identity:}
The Kawamura product, Eq.~\eqref{KawamuraProduct}, is not ``associative'' in the sense of Eq.~\eqref{Associativity} so the fundamental identity is not automatically satisfied.
To investigate whether a subspace exists 
in which the fundamental identity holds, 
consider an $N$-level system.
Let $(E^{abc})_{ijk} \;=\; \delta^a_i\delta^b_j\delta^c_k,$ that is, 
$E^{abc}$ is an $N\times N \times N$ cube with the $abc$ element equal to $1$, and all other elements equal to $0$. Then, the space of all $N$-level hermitian cubes is spanned by the 
following set of basis cubes:
\begin{eqnarray}
E^a & = &  E^{aaa}\;,\quad
F^{ab} = \dfrac{1}{\sqrt{3}}\big(
E^{aab} + E^{aba} + E^{baa}
\big) \;,\cr
J^{abc} & = &  \dfrac{1}{\sqrt{6}}
\big(E^{abc} + E^{bca} + E^{cab} + E^{acb} + E^{bac} + E^{cba}
\big) \;,\cr
K^{abc} & = &  \dfrac{-i}{\sqrt{6}}
\big(E^{abc} + E^{bca} + E^{cab} - E^{acb} - E^{bac} - E^{cba}
\big),\cr
&&
\end{eqnarray}
where different superscripts do not have the same value.
Note that $F^{ab}$ and $F^{ba}$ are different cubes, but
$J^{abc}$ and $K^{abc}$ are respectively symmetric and antisymmetric
under permutations of the superscripts.
There are $N$ cubes of type $E^a$,
$N(N-1)$ cubes of type $F^{ab}$, 
and $N(N-1)(N-2)/6$ cubes each of types $J^{abc}$ and $K^{abc}$.
Of these, cubes of type $K^{abc}$ are purely imaginary while all the other cubes are real. 
As mentioned earlier, $E^a$ and $F^{ab}$ have hermitian matrix analogs, cf.
Eq.~\eqref{MatrixCubeCorrespondence}.
Normalizations are chosen so that
$(A,B) = \delta_{AB}$ for any of pair of the above basis cubes.

Since the space of hermitian cubes is closed under $i$ times the termutator, it endows the space with a ternary algebra~\cite{Abramov2009AlgebrasWT, Bruce2022, Aader2025, Blumenhagen:2025nla}
(aka a Nambu-Lie algebra~\cite{Takhtajan:1993vr}) structure.
For the basis $\{E^a,F^{ab},J^{abc},K^{abc}\}$,
the structure constants
\begin{equation}
i[A,B,C]\;=\;f_{ABCD}D,
\end{equation}
are found to be totally antisymmetric in the four indices, so we will only specify
the ones that cannot be obtained by a permutation of the four cubes.
They are
\begin{eqnarray}
i[J^{abc},J^{abd},J^{acd}] & = &
i[J^{abc},K^{abd},K^{acd}] \; = \;
-\frac{1}{6}K^{bcd}\;,
\cr
i[J^{abc},J^{abd},K^{acd}] & = &
i[K^{abc},K^{abd},K^{acd}] \; = \; 
\frac{1}{6}J^{bcd},
\cr 
i[F^{ab},F^{ac},J^{abc}] 
& = & 
-\frac{1}{3}K^{abc}\;,
\label{TermutationRelations}
\end{eqnarray}
where $a$, $b$, $c$, and $d$ are all different.
Note that $i$ times the termutator of three real cubes, 
or one real cube and two purely imaginary $K^{abc}$s, can only close on the $K^{abc}$s.
Similarly, $i$ times the termutator of three purely imaginary $K^{abc}$s, or one $K^{abc}$ and two real cubes, can only close on the reals.
All other termutators 
that cannot be obtained from those listed here
by permutations of the cubes are zero.  In particular, the $E^{a}$ cubes termute with every other element, so they constitute the ``center'' of the algebra, and this ensures the conservation of $\Tr\,\rho$.
Also, termutators in the first two lines can only exist when $N \ge 4$.

Using the above termutation relations, 
we find that the 4-dimensional subspace spanned by the following set is
a subalgebra on which the fundamental identity is satisfied:
\begin{equation}
\{F^{ab},F^{ac},J^{abc},K^{abc}\}\;.
\end{equation}
Another such set is
\begin{equation}
\{L^{abc},L^{abd},L^{acd},L^{bcd}\},
\label{CubicSpin}
\end{equation}
where 
each $L$ represents either a $J$ or a $K$, with the restriction that one or three of the $L$s are $J$s and the remainder $K$s.
The algebra spanned by Eq.~\eqref{CubicSpin} 
was called the ``cubic spin algebra'' in \cite{Kawamura:2003cw}, in analogy with the spin algebra in canonical QM. 

Both these classes of subalgebras are simple.

The fundamental identity will also hold on direct sums of
these simple 4-dimensional subalgebras when they do not share any indices.
The center consisting of the $E^a$s can also be added to the direct sum.
The dynamics governed by Eq.~\eqref{NambuQM} will make sense on the 
subalgebra constructed in this way.

\noindent
\textit{Dynamics of Triple-Path Interferences:}
This restriction of the dynamics to only a subspace of the hermitian cubes may seem problematic if one wishes 
Eq.~\eqref{NambuQM} to encompass canonical
QM dynamics as well as the additional dynamics of the triple-path interferences. 
However, this would be impossible as long as 
the termutator of any triple of elements with
density matrix analogs is
identically zero, even if the fundamental identity held over the entire space.
This fact could be analogous to ternary Nambu dynamics not necessarily being able to internalize canonical Poisson dynamics~\cite{Takhtajan:1993vr}.
Thus, our strategy is to separate the dynamics of
the elements of the density cube that have density matrix analogs, cf. Eq.~\eqref{MatrixCubeCorrespondence}, and that of the triple-path interference elements so that the two sets do not mix.
The elements that can be mapped to the density matrix can be
evolved according to canonical QM, and the triple-path interferences $J^{abc}$ and $K^{abc}$ evolved via Eq.~\eqref{NambuQM}.

For instance, for $N=3$ consider the subalgebra spanned by
$\{F^{12},F^{13},J^{123},K^{123}\}$ and let $H_1=F^{12}$ and $H_2=F^{13}$.
Let
\begin{equation}
\begin{split}
\rho(t) =
\psi_1(t) J^{123} + \psi_2(t) K^{123}\;,
\end{split}
\end{equation}
where $\psi_{1,2}(t)$ are both real.
Then, Eq.~\eqref{NambuQM} leads to 
\begin{eqnarray}
\dot{\psi}_1 = \frac{1}{3}\psi_2\;,\quad
\dot{\psi}_2 =  -\frac{1}{3}\psi_1\;,
\label{evolutioneq1}
\end{eqnarray}
\noindent
that is, $J^{123}$ and $K^{123}$ oscillate into each other.
If we are to avoid using $F^{ab}$ as the Hamiltonian,
then we must have $N\ge 4$ for ternary dynamics to exist.

For $N=4$, consider the subalgebra spanned by 
$\{J^{123},J^{124},J^{134},K^{234}\}$ and let 
$H_1 = J^{123}$ and  $H_2 = K^{234}$. Let
\begin{equation}
\rho(t) = 
\psi_1(t)J^{124} + \psi_2(t)J^{134}\;,
\end{equation}
where $\psi_{1,2}(t)$ are real.
Then, Eq.~\eqref{NambuQM} leads to
\begin{eqnarray}
\dot{\psi}_{1} = \frac{1}{6}\psi_{2}\;,\quad
\dot{\psi}_{2} = -\frac{1}{6}\psi_{1}\;.
\label{evolutioneq4}
\end{eqnarray}
This describes the oscillation between
$J^{124}$ and $J^{134}$.

Thus, in this approach, the termutation relations lead to
oscillations between pairs of triple-path interferences.
Since we do not have a rule which determines the Hamiltonians $H_1$ and $H_2$, and how they are related to the canonical Hamiltonian for the density matrix, it is unclear what the oscillation frequency should be.
The answer will require a more complete theory from which
the current formalism emerges.

The fact that the formalism of density cubes has an inherent basis dependence could also be an indication that to incorporate triple-path and higher-order interference effects, selection of a preferred basis may be necessary. Though failure of one or more of our cherished notions in canonical QM is to be expected when we try to incorporate a post-quantum phenomenon, we hope that the further development of this work will contribute to a better understanding of its physical implications.

\noindent
\textit{Conclusion \& Discussion:}
We have proposed an equation of motion for density cube theory \cite{Dakic_2014} that leads to triple-path interferences
oscillating among themselves. 
Since it provides a potential description of the time dependence of triple-path interference,
it may have implications for the modeling and interpretation of experiments that aim to place bounds on their sizes~\cite{Sinha:2010, Park:2012, Sawant2014, Kauten_2017, Cotter2017, Sadana2022, Namdar2023, Huber2022}.
In particular, the experimental probes of quantum properties of gravity are just being developed \cite{Bose_2025}, and our approach could be used to model experiments that aim to use massive quantum probes for study of possible intrinsic triple and higher order quantum interference in the realm of quantum gravity~\cite{Berglund:2023vrm, Berglund:2025yoy}.

The totally antisymmetric Nambu bracket, Eq.~\eqref{NambuBracket}, can
be defined for any number $N$ of objects, and Nambu dynamics, Eq.~\eqref{NambuCM}, extended to that with $N-1$ Hamiltonians.
We envision ``quantizing'' the $N$-ary Nambu dynamics
to obtain a dynamical theory of $N$-path interferences, following the
procedure we developed for the ternary case in this letter.
Indeed, the $N$-ary product has already been defined by Kawamura in \cite{Kawamura:2005ic}, Appendix A, as a direct generalization of Eq.~\eqref{KawamuraProduct} in a way that preserves the ``hermiticity'' of the $N$-ary commutator.
Thus, we may be able to extend the formalism order by order.

At each order in the extension, new degrees of freedom will be introduced, just as the density cube $\rho_{ijk}$ has extra parameters beyond those in the density matrix $\rho_{ij}$.
Such new degrees of freedom may appear naturally in the context of quantum gravity.
Note that in all quantum theories in which spacetime is treated as a
non-quantum background, the Born rule for computing quantum probabilities is fixed.
There are many arguments why the Born rule is robust in quantum mechanics and quantum field theory (for a summary, see the discussion in~\cite{Berglund:2023vrm, Berglund:2025yoy}),
but these arguments always rely on assumptions which are incompatible with the very nature of gravity, such as
its dynamically causal structure, as well as the fact that gravity cannot be screened, and that energy, i.e., the Hamiltonian, is the gravitational charge, or that gravity has an intrinsic geometric entropy, unlike other fundamental forces.
Thus the Born rule cannot be assumed to be valid in quantum gravity, which is a theory of quantum spacetime~\cite{Berglund:2023vrm, Berglund:2025yoy}. Also the experimental validity of the Born rule has not been established in quantum gravity~\cite{Bose_2025} and in order to check its validity one needs a formalism, such as the one presented in this paper, which allows for the phenomena that go beyond the implications of the quadratic Born rule, such as intrinsic triple-path interference. In principle, one can argue from many points of view that the Born rule is dynamical in quantum gravity, in complete analogy with the dynamical nature of spacetime in general relativity, and in this case one speaks of ``gravitization of quantum theory''~\cite{hubsch2024, Hubsch:2026lot}. In that context, new observables such as density cubes, are natural and, one might argue, unavoidable.

Finally, we note that the motivation for exploring triple and higher-order interference has come from both quantum foundations \cite{Barnum_2014, Lee_2016, Ududec_2010}, where the approach is often operational\cite{Hardy2001quantum}, and quantum gravity \cite{Berglund:2023vrm, hubsch2024}, where dynamical models are explicitly written down. Our work, since it connects density cube theory\cite{Dakic_2014} with studies inspired by quantum gravity\cite{Awata:1999dz, Yoneya:2021lyf}, could help bridge the gap between the two fields.

\noindent
{\bf Acknowledgments:}
We thank  P.~Berglund, A.~Geraci, T.~H\"{u}bsch, and D.~Mattingly for discussions. DM thanks the Perimeter Institute for hospitality. This work is supported in part by the U.S. Department of Energy under contract DE-SC0020262.

\bibliography{DensityCube}
\bibliographystyle{apsrev4-2}

\end{document}